\documentclass[preprint,superscriptaddress,amsmath,amssymb,aps]{revtex4-1}

\usepackage{graphicx}
\usepackage{dcolumn}
\usepackage{bm}

\begin{document}


\title{Unravelling the Local Crystallographic Structure of Ferromagnetic $\gamma$'-Ga$_y$Fe$_{4-y}$N Nanocrystals Embedded in GaN}


\author{A. Navarro-Quezada}
\email{andrea.navarro-quezada@jku.at}
\affiliation{Institute of Semiconductor and Solid-State Physics, Johannes Kepler University Linz, Altenberger Str. 69, 4040 Linz, Austria}

\author{K. Gas}
\affiliation{Institute of Physics, Polish Academy of Sciences, Aleja Lotnikow 32/46, PL-02668 Warsaw, Poland}

\author{A. Spindlberger}
\affiliation{Institute of Semiconductor and Solid-State Physics, Johannes Kepler University Linz, Altenberger Str. 69, 4040 Linz, Austria}

\author{F. Karimi}
\affiliation{SOLEIL Synchrotron, Gif-sur-Yvette, France}

\author{M. Sawicki}
\affiliation{Institute of Physics, Polish Academy of Sciences, Aleja Lotnikow 32/46, PL-02668 Warsaw, Poland}

\author{G. Ciatto}
\affiliation{SOLEIL Synchrotron, Gif-sur-Yvette, France}

\author{A. Bonanni}
\affiliation{Institute of Semiconductor and Solid-State Physics, Johannes Kepler University Linz, Altenberger Str. 69, 4040 Linz, Austria}


\date{\today}

\begin{abstract}
In the Fe-doped GaN phase-separated magnetic semiconductor Ga$\delta$FeN, the presence of embedded $\gamma$’-Ga$_y$Fe$_{4-y}$N nanocrystals determines the magnetic properties of the system. Here, through a combination of anomalous x-ray diffraction and diffraction anomalous fine structure, the local structure of Ga in self-assembled face-centered cubic (fcc) $\gamma$’-Ga$_y$Fe$_{4-y}$N nanocrystals embedded in wurtzite GaN thin layers is investigated in order to shed light onto the correlation between fabrication parameters, local structural arrangement and overall magnetic properties of the material system. It is found, that by adjusting the growth parameters and thus, the crystallographic surroundings, the Ga atoms can be induced to incorporate into 3$c$ positions at the faces of the fcc crystal lattice, reaching a maximum occupancy of 30\%. The magnetic response of the embedded nanocrystals is ferromagnetic with Curie temperature increasing from 450\,K to 500\,K with the Ga occupation. These results demonstrate the outstanding potential of the employed experimental protocol for unravelling the local structure of magnetic multi-phase systems, even when embedded in a matrix containing the same element under investigation.
\end{abstract}


\maketitle

\section{Introduction}

The incorporation of Fe ions into the technologically relevant GaN fabricated by metal-organic vapour phase epitaxy above a concentration of 0.4\% cations leads to the self-assembly of embedded Fe$_y$N nanocrystals (NCs) with either ferromagnetic  or antiferromagnetic properties imposed by the growth conditions and by the actual incorporation of the magnetic ions into the wurtzite lattice\,\cite{Bonanni:2007_PRB, Navarro:2010_PRB}. These Fe$_y$N nanocrystals are expected to provide an efficient, robust and energy-saving platform in particular for magnetic data storage\,\cite{Coey:1999_JMMM} and for the generation of polarized spin currents. Further applications include spin current injection into the semiconductor host crystal through spin pumping, and spin current detection by inverse spin Hall effect\,\cite{Chen:2013_NatCom}. Moreover, these semiconductor/NCs hybrid structures offer the possibility to study frustrated magnetic systems and spin glass states\,\cite{Tupan:2020_materials}.\\

Particularly relevant is the $\gamma$’-Fe$_4$N phase, which has a nearly perfect minority spin polarization and is ferromagnetic up to temperatures as high as 765\,K\,\cite{Shirane:1962_PR, Kokado:2006_PRB}, making it suitable for high-density magnetic storage\,\cite{Tsunoda:2010_APEX}. Investigations on the effect of the presence of Ga in $\gamma$’-Fe$_4$N powders are also available\,\cite{Houben:2009_ChemMat, Burghaus:2010_IC, Burghaus:2011_JSSC}. When Ga is incorporated into the $\gamma$’-Fe$_4$N perovskite Pm$\bar{3}$m face-centered cubic (fcc) structure, the Ga atoms can occupy the 1$a$ (corner) or the 3$c$ (faces) Wyckoff positions. It is reported that, Ga tends to occupy the energetically favorable 1$a$ Wyckoff positions, resulting in a maximum occupancy of 90\% Ga and 10\% Fe\,\cite{Burghaus:2010_IC}. The magnetic response of the compound is largely modified through the incorporation of Ga and it ranges from ferromagnetic for $0\,<\,y\,<\,0.25$ to weakly antiferromagnetic for $y\,>0.25$, accompanied by a reduction of the magnetic moment and of the coercive field\,\cite{Burghaus:2011_JSSC, Houben:2009_ChemMat}.\\

In this frame, the growth of ordered arrays of magnetic $\gamma$’-Ga$_y$Fe$_{4-y}$N NCs embedded in GaN\,\cite{Navarro:2012_APL, Grois:2014_Nanotech, Navarro:2019_crystals}, whose size, shape and density can be tuned on demand through the amount of Ga provided during deposition, becomes particularly appealing.  By controlling the growth conditions, the incorporation of Ga atoms into the 1$a$ Wyckoff positions of the fcc $\gamma$’-Fe$_4$N lattice is expected to be promoted, altering the magnetic characteristics of the NCs and therefore, the overall magnetic properties of the layers. In order to implement these nano-objects in ferromagnetic\,\cite{Ando:2011_NatMat, Chen:2013_NatCom}, as well as in antiferromagnetic spintronic devices\,\cite{Wadley:2016_Science, Gomonay:2014_LTP}, it is imperative to identify and quantify the local arrangement of the Ga atoms in the $\gamma$’-Ga$_y$Fe$_{4-y}$N NCs embedded in the GaN host. However, the surrounding GaN matrix challenges the detection -- \textit{e.g.} by conventional element-selective x-ray absorption spectroscopy -- of Ga atoms incorporated into the NCs.\\

Here, through a combination of anomalous x-ray diffraction (AXD) and diffraction anomalous fine structure (DAFS)\,\cite{Maurizio:2020_ASS, Stragier:1992_PRL, Pickering:1993_JACS}, the local-structure of self-assembled $\gamma$’-Ga$_y$Fe$_{4-y}$N nanocrystals embedded in GaN thin layers is investigated, in order to shed light onto the correlation between fabrication parameters, local crystallographic arrangement and overall magnetic properties of the material system. It is found, that due to the conditions imposed by the growth parameters and by the resulting crystallographic environment, the Ga atoms tend to occupy the 3$c$ sites of the fcc structure, while the magnetic characteristics of the NCs -- magnetization and Curie temperature ($T_\mathrm{C}$) -- display an inverse correlation with the amount of Ga atoms occupying 3$c$ sites in the NCs.
The present results demonstrate the outstanding potential of the DAFS method for unravelling the local structure of similar magnetic phase-separated systems even when the NCs and the surrounding matrix contain the same absorber under investigation, though in a different crystallographic landscape.\\

\section{Results and Discussion}
\begin{figure}
	\centering
	\includegraphics[width=13 cm]{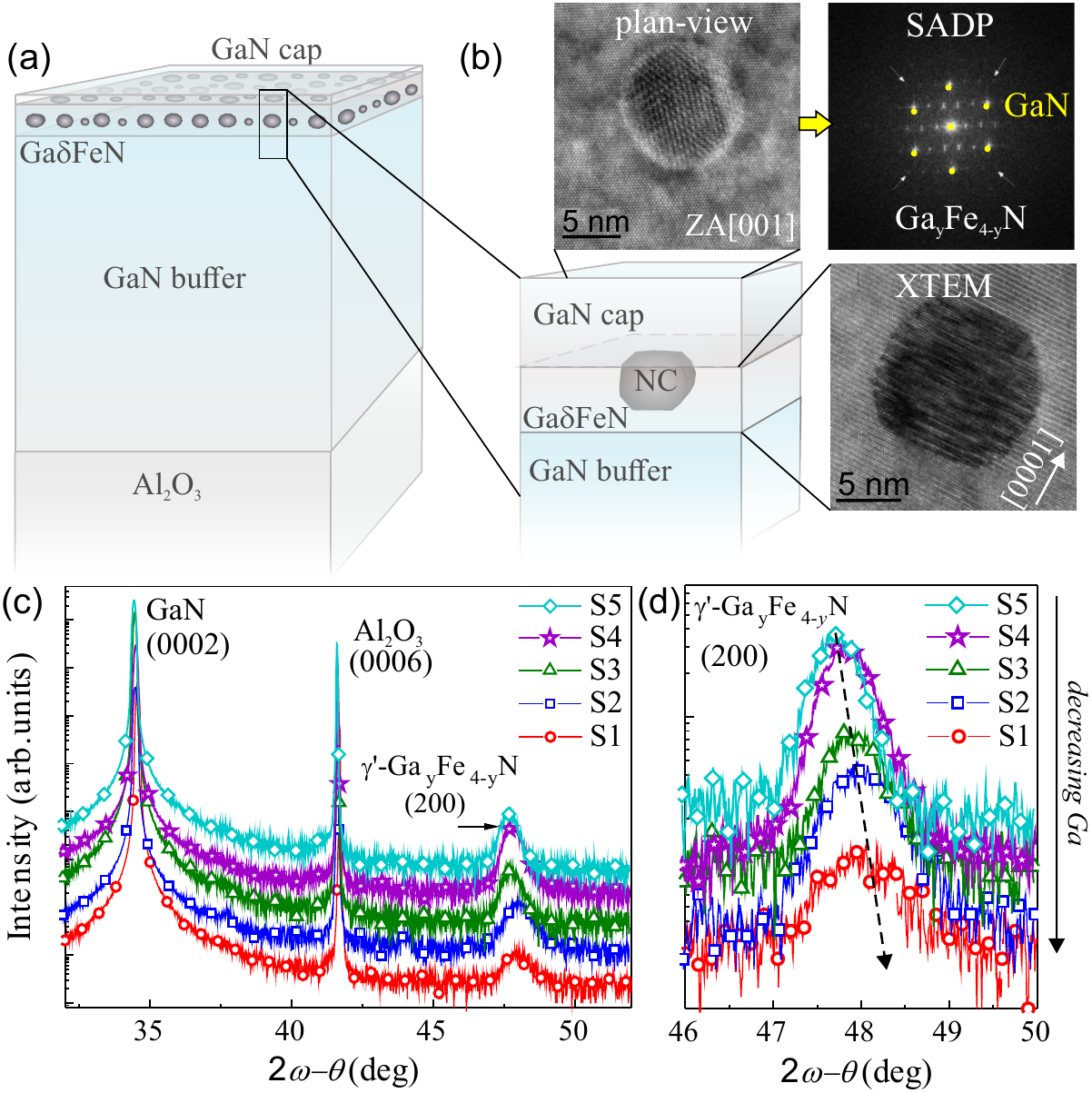}
	\caption{(a) Schematic structure of the phase-separated Ga$\delta$FeN layers. (b) Detail of a nanocrystal and its surroundings with corresponding plan-view TEM, SADP and cross-section TEM (XTEM) image. (c) Radial 2$\theta$-$\omega$ HRXRD scans of the investigated samples. (d) Detail of the scan in (c) showing the shift of the (200) $\gamma$'-Ga$_y$Fe$_{4-y}$N diffraction peak to larger angles, \textit{i.e} smaller lattice parameter along the $c$-growth direction, with decreasing normalized Ga flow.}
	\label{fig:gafen}
\end{figure}

The investigated series of GaN thin layers with embedded $\gamma$'-Ga$_y$Fe$_{4-y}$N NCs are fabricated by metal organic vapour phase epitaxy according to the procedure detailed in Refs.\,\cite{Navarro:2019_crystals, Navarro:2012_APL} by varying the Ga supply over the samples' series in order to adjust the incorporation of Ga atoms into the fcc $\gamma$’-Fe$_4$N structure of the NCs. The Fe-doped GaN thin layers, denoted from here on as Ga$\delta$FeN (Section\,\ref{exp}), are grown on a 1\,$\mu$m thick GaN buffer, deposited onto a $c$-plane sapphire (Al$_2$O$_3$) substrate. The samples are covered with a GaN capping layer with thickness ranging between 40\,nm to 300\,nm to prevent the formation of $\alpha$-Fe NCs at the sample surface\,\cite{Li:2008_JCG, Navarro:2019_PRB}. A schematic representation of the sample structure is presented in Fig.\,\ref{fig:gafen}\,(a). The nanocrystals’ size varies depending on the amount of Ga provided during fabrication and is limited to (17$\pm$8)\,nm along the $c$-growth direction, while it ranges between (50$\pm$20)\,nm and (20$\pm$8)\,nm in the layer plane\,\cite{Navarro:2019_crystals}, as exemplified in the plan-view and cross section high-resolution transmission electron microscopy (HRTEM) images of a nanocrystal shown in Figure\,\ref{fig:gafen}\,(b). The selective area diffraction pattern (SADP) of the NCs in the plan-view image allows identifying the GaN matrix and the fcc $\gamma$'-Ga$_y$Fe$_{4-y}$N phase of the NCs from the diffraction spots. Moreover, in the SADP taken along the [001] zone-axis (ZA), the in-plane epitaxial relation $\langle 110 \rangle_\mathrm{NC}$$\parallel$$\langle 1100\rangle_\mathrm{GaN}$ between the nanocrystal and the GaN matrix is evidenced. The nanocrystals density is also tuneable by the amount of Ga provided during growth and is listed in Table\,\ref{tab:param} as a function of the normalized Ga-flow. This parameter is obtained by multiplying the trimethyl-gallium (TMGa) source flow by the time over which the TMGa source is kept open during the growth of the Ga$\delta$FeN layer and of the subsequent GaN capping film, as the incorporation by diffusion of Ga into the NCs during the fabrication of the capping layer cannot be ruled out.
\begin{table}
	\centering
	\caption{Investigated samples: growth parameters, lattice parameter along the $c$-growth direction calculated from HRXRD, NCs density\,\cite{Navarro:2019_crystals}, and fit parameters extracted from DAFS.}
	\begin{tabular}{ccccccc}
		\hline
		Sample & Normalized Ga-flow& Cap thickness & Lattice parameter & NC density & $(A-B)/A$ & $y^{3c}$\\
		Number & (arb. units) & (nm) & $\pm 0.005$\,($\AA$) & 10$^{9}$ (NC/cm$^2$) & & \\
		\hline
		S1 & 104  & 40 & 3.789 & 3.9 &  0.07 & 0.10 \\
		S2 & 126  & 50 & 3.786 & 7.1 & 0.13 & 0.18 \\
		S3 & 218  & 140 & 3.797 & 9.8 & 0.19 & 0.26 \\
		S4 & 312  & 230 & 3.798 & 10.2 & 0.22 & 0.30 \\
		S5 & 350  & 300 & 3.806 & 15.0 & 0.23 & 0.31 \\
		\hline
	\end{tabular}
	\label{tab:param}
\end{table}

\subsection{Structure}
The 2$\theta$-$\omega$ high-resolution x-ray diffraction (HRXRD) scans depicted in Fig.\,\ref{fig:gafen}\,(c) for the Ga$\delta$FeN layers show that besides the diffraction peaks of the crystal planes for GaN (0002) and for the underlying Al$_2$O$_3$ (0006) substrate, only an additional diffraction peak assigned to the (200) diffraction planes of the $\gamma$’-Ga$_y$Fe$_{4-y}$N embedded nanocrystals emerges. The $\langle100\rangle$ directions of the NCs lie parallel to the [0001] $c$-growth direction of the wurtzite matrix. The shift of the diffraction peak of the NCs towards larger diffraction angles with decreasing normalized Ga-flow observed in Fig.\,\ref{fig:gafen}\,(d)\,\cite{Navarro:2012_APL, Navarro:2019_crystals}, points at an increase of the lattice parameter of the NCs along the $c$-growth direction that is tentatively assigned to the inclusion of Ga into the $\gamma$’-Fe$_4$N crystal structure\,\cite{Burghaus:2010_IC}. The calculated lattice parameters from the HRXRD are given in Table\,\ref{tab:param}.

\begin{figure}
	\centering
	\includegraphics[width=15 cm]{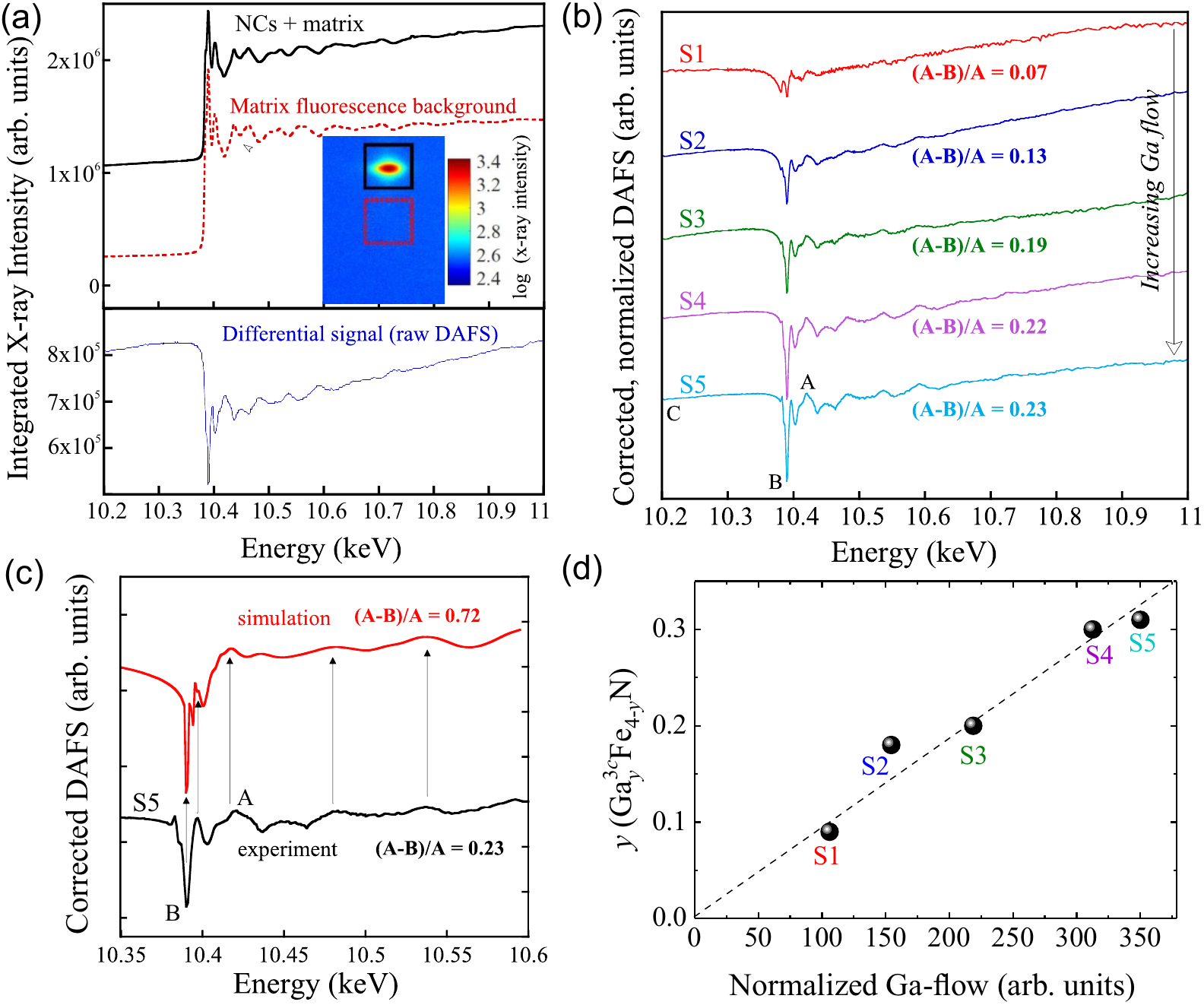}
	\caption{(a) Top: Ga $K$-edge DAFS for sample S5 acquired at the (002) diffraction plane of the NCs including the fluorescence from the matrix (---) and the pure matrix fluorescence signal (- -). Inset: intensity map and ROIs from where the two spectra have been obtained. Bottom: differential spectrum obtained from the two spectra shown in (a). (b) Corrected, normalized DAFS for all samples. The Ga-flow during the fabrication increases from top to bottom. (c) Comparison between DAFS simulation (top) and experimental data of S5 (bottom) with Ga occupying 3$c$ sites. (d) Amount of Ga $y$ in 3$c$ sites in the $\gamma$’-Ga$^{3c}_y$Fe$_{4-y}$N NCs embedded in GaN plotted as a function of the normalized Ga-flow, obtained from the fabrication parameters.}
	\label{fig:dafs}
\end{figure}

In order to confirm the inclusion of Ga atoms into the NCs, AXD and DAFS are carried out at the SIRIUS beamline at SOLEIL at the Ga $K$-edge in the energy range from 10.2\,keV to 11\,keV on the (200) diffraction plane of the NCs. In particular, the combination of the two techniques allows addressing this issue directly since, after the raw signal has been properly corrected according to the procedure detailed here below, no anomalous variation at the Ga absorption edge is expected for a phase such as the Ga-free ferromagnetic $\gamma$’Fe$_4$N one. The main challenge, which can be overcome by employing DAFS, is the presence of Ga atoms in the surrounding matrix and in the capping layer. In order to subtract the substantial x-ray fluorescence background generated by the host crystal, two regions of interest (ROIs) have been selected during the DAFS measurements shown in the inset to Fig.\,\ref{fig:dafs}\,(a): a ROI around the (002) NC diffraction peak (solid line square), and another one (of the same area and close to it) containing solely the fluorescence of the Ga in the surrounding GaN host crystal (dashed line square). The fluorescence from the GaN matrix is then subtracted from the spectra obtained from the (002) NC diffraction, leaving the contribution of Ga to the NCs (002) diffraction only. The raw Ga $K$-edge DAFS signal of the NCs including the fluorescence background due to the GaN surroundings in sample S5 is exemplarily shown in Fig.\,\ref{fig:dafs}\,(a) as a solid line, while the dotted line spectrum in the same figure represents the fluorescence background only. The differential signal presented in Fig.\,\ref{fig:dafs}\,(a) points at the inclusion of Ga atoms into the NCs. 

A detailed analysis of the DAFS spectra has been performed by applying the following corrections to the raw signal in order to recover a signal proportional to the square of the structure factor: absorption correction from air, by the capping layer and by the Ga$\delta$FeN layer, Lorentz factor, polarization, filters, detector efficiency, and geometrical factor\,\cite{Prioetti:1999_PRB}. The data are normalized to the same pre-edge intensity for all samples. In Fig.\,\ref{fig:dafs}\,(b), the corrected, normalized differential DAFS spectra for all samples with increasing Ga-flow from top to bottom are presented. The spectra are shifted vertically for clarity. The variable $(A-B)/A$, with the diffracted intensities $A$ and $B$ marked in the spectra in Fig.\,\ref{fig:dafs}\,(b), quantifies the absorption edge and, therefore, the contribution of the Ga atoms to the structure factor of the reflection. The parameter $A$ corresponds to the normalized and corrected diffracted intensity after the edge, lying around an energy value of 10.4\,keV for all samples, while $B$ is the diffracted intensity at the edge, \textit{i.e.} at the minimum of the DAFS spectrum, located around 10.39\,keV for all samples. Therefore, the difference $A-B$ gives the size of the anomalous diffraction at the Ga $K$-edge. Since the diffracted intensity may vary from sample to sample, this difference is normalized for comparison. In this way, $(A-B)/A$ is proportional to the number of Ga atoms which participate in producing the observed reflection. For Ga-free nanocrystals, $(A-B)/A$ would be 0. Thus, this parameter quantifies how many of the Wyckoff sites -- either 1$a$ or 3$c$ -- in the Pm$\bar{3}$m fcc structure of $\gamma$'-Fe$_4$N are occupied by Ga atoms.

Based on previous literature reports\,\cite{Houben:2009_ChemMat, Burghaus:2010_IC, Burghaus:2011_JSSC}, the DAFS spectra have been first simulated in the region of the Ga $K$ absorption edge assuming that Ga occupies exclusively the 1$a$ sites in the fcc structure. However, with this approach a maximum occupancy of 100\% of 1$a$ sites by Ga in samples S5 and S4 is obtained, which is in contrast to the magnetic properties of the samples discussed in the following section and to the maximal occupation of 90\% achieved experimentally by Burghaus \textit{et al.} for $\gamma$’-Ga$_y$Fe$_{4-y}$N powders\,\cite{Houben:2009_ChemMat,
	Burghaus:2010_IC, Burghaus:2011_JSSC}.

Therefore, the occupation of 3$c$ sites by Ga atoms has been also considered. Assuming that all Ga atoms occupy the 3$c$ sites of the fcc structure, the simulated spectra shown in Fig.\,\ref{fig:dafs}\,(c) is obtained,  yielding $(A-B)/A=$\,0.72. For comparison, the experimental spectrum of S5, also presented in Fig.\,\ref{fig:dafs}\,(c), has a value of $(A-B)/A=$\,0.23. The spectral features (marked with arrows) are well reproduced by the simulation, supporting the assumption of the incorporation of Ga into 3$c$ sites. The values of $(A-B)/A\leq0.23$ indicate that a significant amount of 3$c$ sites is not
occupied by Ga atoms or that there is a number of Ga-free NCs, \textit{i.e.} having the pure $\gamma$-Fe$_4$N phase. The magnitudes of the $(A-B)/A$ parameter obtained for each sample and summarized in Table\,\ref{tab:param}, allow estimating the amount $y$ of Ga occupying 3$c$ sites in the NCs. This is plotted as a function of the normalized Ga-flow in Fig.\,\ref{fig:dafs}\,(d), showing a linear correlation between the two.
\begin{figure}
	\centering
	\includegraphics[width= 15 cm]{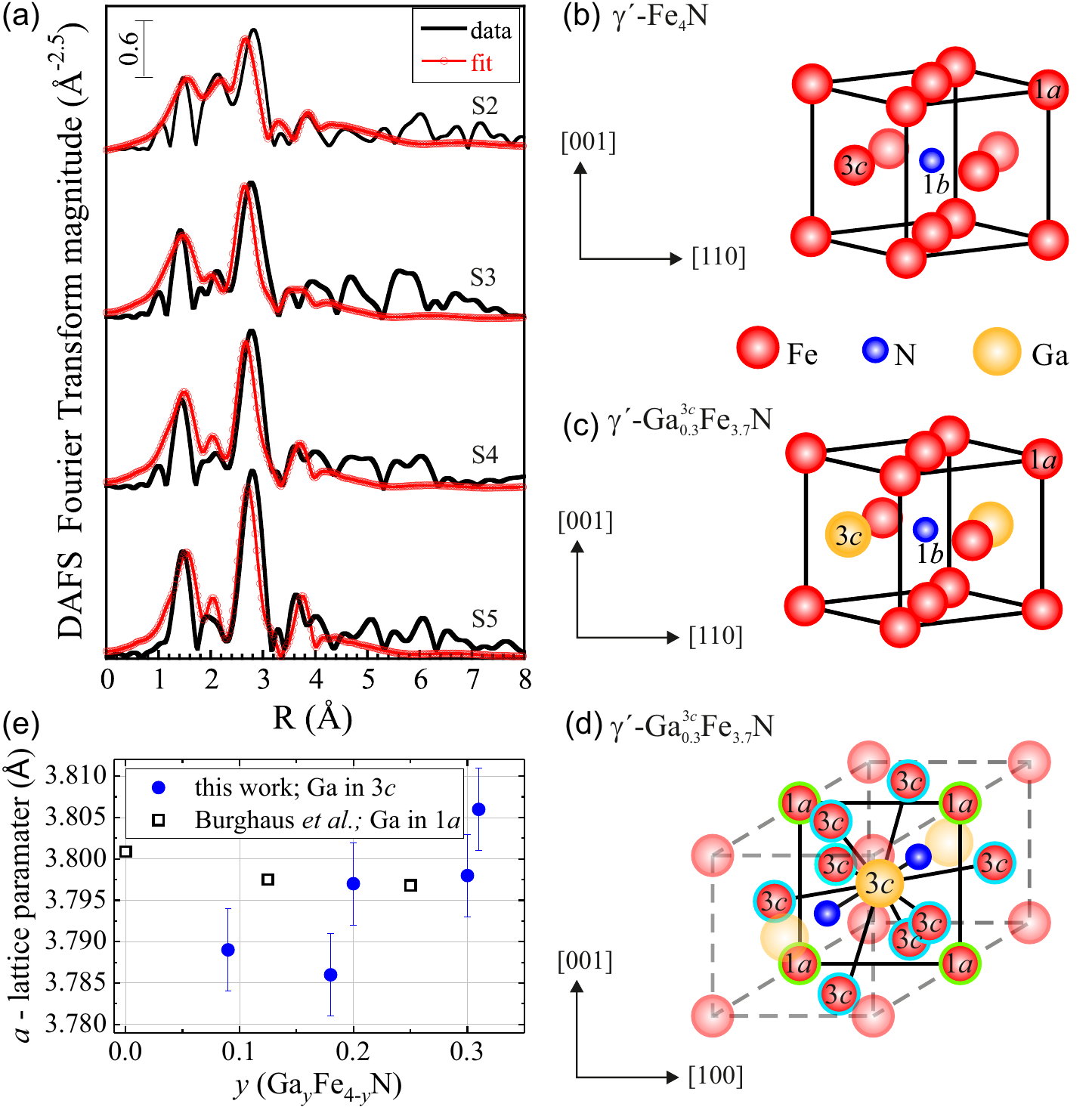}
	\caption{(a) Fourier transform of the DAFS spectra and relative fits. (b) and (c) Pm$\bar{3}$m unit cell of $\gamma$’-Fe$_4$N and $\gamma$'-Ga$^{3c}_{0.3}$Fe$_{3.7}$N (samples S5 and S4), respectively, showing the different occupation sites. (d) Next nearest neighbors ($nnn$) of a Ga atom in a 3$c$ occupation site: four Fe atoms located at the vertices (1$a$) and eight Ga atoms located at the faces (3$c$). The nearest neighbors are the N atoms at the center of the unit cell.(e) Average lattice parameter along the $c$-growth direction of the NCs obtained from the (200) peak position in the 2$\theta$-$\omega$ HRXRD scan as a function of the calculated $y$. The values reported by Burghaus \textit{et al.}\,\cite{Burghaus:2010_IC} for $\gamma$’-Ga$_y$Fe$_{4-y}$N, where Ga occupies exclusively 1$a$ sites are included for comparison.}
	\label{fig:dafs2}
\end{figure}

The occupation of the 3$c$ sites by Ga atoms is further confirmed by the detailed analysis of the Fourier transform (FT) of the corrected DAFS oscillations presented in Fig.\,\ref{fig:dafs2}\,(a). A peak at the apparent distance $R$\,$\approx$\,1.5\,$\AA$ in the first Ga-N correlation shell, corresponding to about $\approx$\,2.0\,$\AA$, is observed for all samples, pointing at Ga in the 3$c$ positions. No N first neighbors at a similar distance are expected if Ga occupies 1$a$ sites. Hence, there is no evidence of a significant amount of Ga atoms occupying 1$a$ positions. The quantitative analysis of the data, which has been performed following the approach described in Refs.\,\cite{Prioetti:1999_PRB, Ciatto:2005_PRB, Ciatto:2018_JAP}, and the resulting structural parameters are collected in Table\,\ref{tab2}. For sample S1, the weakness of the fine structure signal does not allow a reliable quantitative analysis. 

The unit cells of $\gamma$'-Fe$_4$N and $\gamma$'-Ga$_{0.3}^{3c}$Fe$_{3.7}$N are presented in Figs.\,\ref{fig:dafs2}\,(b) and (c) and the occupation sites and coordination distances in the Pm$\bar{3}$m crystal structure of $\gamma$'-Ga$_y$Fe$_{4-y}$N compounds are listed in Table\,\ref{tab3}. When a Ga atom occupies the 3$c$ Wyckoff position in the Pm$\bar{3}$m crystal sructure of $\gamma$'-Fe$_4$N, the nearest neighbors will be the N atoms at the 1$b$ position separated by 1.89\,$\AA$ from the Ga reference position. Further, in this case, the 3$c$ Ga atom has 12 next nearest neighbors ($nnn$) at a distance of 2.68\,$\AA$: four Fe atoms occupying 1$a$ positions (at the vertices of the structure), and eight Fe(Ga) atoms in 3$c$ positions, as schematically shown in Fig.\,\ref{fig:dafs2}\,(d) for a $\gamma$'-Ga$_{0.3}^{3c}$Fe$_{3.7}$N sample. Therefore, in the first atomic shell of Ga, an unique Ga-N distance $R_{\rm Ga-N}$ has been introduced as a variable in the fits, together with the relative Debye-Waller factor $\sigma^2_{\rm Ga-N}$. In the second shell, $R_{\rm Ga-Fe}$ is the distance between Ga at the 3$c$ position and the four Fe $nnn$ at the vertices of the cell, while $R_{\rm Ga-Ga}$ is the one to the eight Fe(Ga) $nnn$ on the faces, considering Fe and Ga as scattering centers when constructing the theoretical paths used in the fit. 

The results reported in Table\,\ref{tab2} show that, for most samples, the ideal crystallographic structure is distorted and the two groups of $nnn$ are found at positions slightly different from the natural ones. The analysis of the shape of the FT and the value of the Debye-Waller factors hint at a global reduction in the amplitude of the signal with reducing the normalized Ga flow. The lineshapes of the different FT spectra are similar with the only exception of sample S2,  where a redistribution of the distances in the second shell is found. The table includes also the extracted
value for the second shell Debye-Waller factors $\sigma^2_{\rm Ga-Fe}$ and $\sigma^2_{\rm Ga-Ga}$.

\begin{table}
	\centering
	\caption{Structural parameters extracted from the quantitative DAFS analysis. $R$-range of the fits = [0.6 - 4.3] $\rm \AA$.}
	\begin{tabular}{ccccccc}
		\hline
		Sample & $R_{\rm Ga-N}$ (${\rm \AA}$) & $R_{\rm Ga-Fe}$ (${\rm \AA}$) & $R_{\rm Ga-Ga}$ (${\rm \AA}$) & $\sigma^2_{\rm Ga-N}$ ($10^{-3}{\rm \AA}^2$) & $\sigma^2_{\rm Ga-Fe}$ ($10^{-3}{\rm \AA}^2)$ & $\sigma^2_{\rm Ga-Ga}$ ($10^{-3}{\rm \AA}^2$)\\
		\hline
		S2  & 1.983 $\pm$ 0.033 & 2.856 $\pm$ 0.127 & 2.799 $\pm$ 0.083 & 0.0115 $\pm$ 0.0000 & 0.0151 $\pm$ 0.0027 & 0.0135 $\pm$ 0.0024\\
		S3  & 1.956 $\pm$ 0.026 & 2.718 $\pm$ 0.025 & 2.841 $\pm$ 0.015 & 0.0085 $\pm$ 0.0057 & 0.0067 $\pm$ 0.0029 & 0.0060 $\pm$ 0.0026\\
		S4  & 1.954 $\pm$ 0.015 & 2.711 $\pm$ 0.020 & 2.838 $\pm$ 0.013 & 0.0036 $\pm$ 0.0031 & 0.0041 $\pm$ 0.0020 & 0.0036 $\pm$ 0.0018\\
		S5  & 1.961 $\pm$ 0.013 & 2.714 $\pm$ 0.025 & 2.843 $\pm$ 0.013 & 0.0019 $\pm$ 0.0022 & 0.0037 $\pm$ 0.0017 & 0.0033 $\pm$ 0.0015\\      	
		\hline
	\end{tabular}
	\label{tab2}
\end{table}

\begin{table}
	\centering
	\caption{Literature values of the occupation sites and coordination distances in the Pm$\bar{3}$m fcc crystal structure of $\gamma$'-Ga$_y$Fe$_{4-y}$N compounds.}
	\begin{tabular}{cccccc}
		\hline
		Compound & Wyckoff position & Occupation &\multicolumn{3}{c}{Distance ($\AA$)}\\
		& & & 1$a$ & 1$b$ & 3$c$\\ 
		\hline
		$\gamma$'-Fe$_4$N\,\cite{Jacobs:1995_JAC} & 1$a$  & Fe & 3.79 & 3.28 & 2.68\\
		& 1$b$  & N &  3.28 & 3.79 & 1.89\\
		& 3$c$ & Fe & 2.68 & 1.89 & 2.68\\
		$\gamma$'-GaFe$_3$N\,\cite{Burghaus:2010_IC} & 1$a$  & Ga(0.9),Fe(0.1) & 3.80 & 3.28 & 2.68\\
		& 1$b$  & N &  3.29 & 3.80 & 1.90\\
		& 3$c$ & Fe & 2.68 & 1.90 & 2.68\\
		\hline
	\end{tabular}
	\label{tab3}
\end{table}
The change in the lattice parameter $a$ along the $c$-growth direction of the NCs, as obtained from the HRXRD spectra as a function of the Ga content $y$ is shown in Fig.\,\ref{fig:dafs2}\,(e), where the NCs in the layers fabricated with low normalized Ga-flow, \textit{i.e.} lowest $y$, display also the smallest lattice parameter. The obtained lattice parameters differ significantly from those obtained by Burghaus \textit{et al.} for $\gamma$'-Ga$_y$Fe$_{4-y}$N powders, where Ga occupies exclusively the 1$a$ sites. It was calculated, that the lattice parameter of $\gamma$'-GaFe$_3$N should be larger by about 2\% -- if Ga occupies the 3$c$ sites -- with respect to the one found if Ga occupies only 1$a$ sites\,\cite{Burghaus:2010_IC}. In the NCs considered here, the lattice parameter for the layers with $y^{3c} > 20$\% is comparable or larger than the one reported by Burghaus \textit{et al.}, as observed in Fig.\,\ref{fig:dafs2}\,(e).

\subsection{Magnetic properties}
The $\gamma$'-GaFe$_3$N ternary system is paramagnetic in the temperature range between 20\,K and 300\,K and weakly antiferromagnetic below 20\,K. Interestingly, a local ferromagnetic ordering of Fe atoms building a 13 atoms cluster when 90\% of the 1$a$ sites are occupied by Ga atoms has been reported for $\gamma$’-GaFe$_3$N powders\,\cite{Burghaus:2011_JSSC}. The magnetic response of phase-separated GaN thin layers containing Fe$_y$N embedded nanocrystals consist of two contributions: (i) a paramagnetic Brillouin-like one dominant below 50\,K, attributed to Fe$^{3+}$ ions with $S = 5/2$ substituting for Ga cations in the matrix, and (ii) a component relevant above 50\,K and associated with the ferromagnetic NCs embedded in the layers\,\cite{Navarro:2010_PRB, Gas:2019_MST, Navarro:2020_materials}. In films where antiferromagnetic $\zeta$-Fe$_2$N NCs are formed, the presence of an additional linear non-saturating component was observed\,\cite{Navarro:2010_PRB}.
\begin{figure}
	\centering
	\includegraphics[width=17 cm]{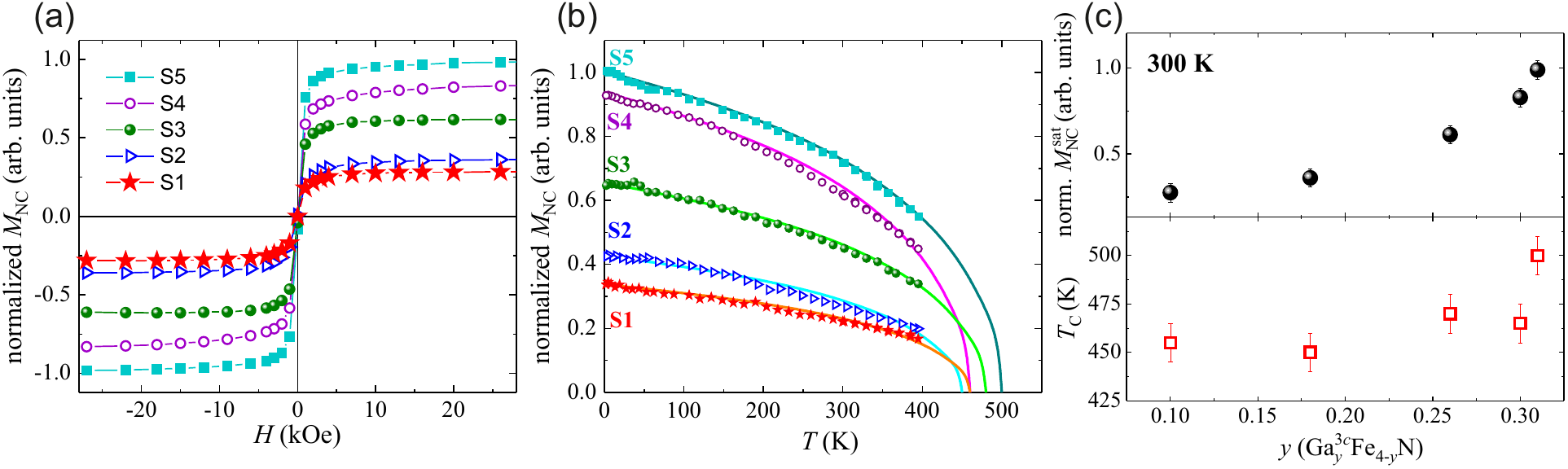}
	\caption{(a) Normalized $M_\mathrm{NC}$ for all samples as a function of magnetic field $H$. (b) Temperature $T$ dependence of the normalized $M_{NC}$ acquired up to 390\,K (symbols). Lines: Langevin $M(T)$ fits. (c) Normalized $M_{NC}^{\mathrm{sat}}$ and $T_\mathrm{C}$ as a function of the amount of Ga $y$ in 3$c$ sites.}
	\label{fig:mag}
\end{figure}

In the samples considered here, the total magnetic moment of the Ga$\delta$FeN layers consists of the two components mentioned above, which can be separated by employing the procedure detailed in Refs.\,\cite{Navarro:2010_PRB,Navarro:2020_materials}. The amount $x_{\mathrm{Fe3+}}$ of Fe$^{3+}$ ions contributing to the Brillouin-like paramagnetism is estimated by considering that the volume of the film nominally containg Fe ions consists of the Ga$\delta$FeN and of the capping layer, where Fe diffusion is expected to take place. For all samples, $x_{\mathrm{Fe3+}}$ \,=\,(0.1$\pm$0.02)\% is obtained, a value well below the established solubility limit of $x_{\mathrm{Fe}}$\,=\,0.4\% for Fe in GaN at the growth conditions considered here\,\cite{Bonanni:2007_PRB, Navarro:2010_PRB}. Since the integrated intensity from the diffraction peak obtained in DAFS is proportional to the volume fraction of the NCs in the samples, the normalized magnetization $M_\mathrm{NC}$ of the nanocrystals is obtained by dividing the magnetic moment of the NCs at 300\,K, after the substraction of the paramagnetic component, by the diffracted intensity. The normalized $M_\mathrm{NC}$ as a function of the field applied in the in-plane configuration ($H\perp$\,[0001], $c$-growth direction) for all samples is presented in Fig.\,\ref{fig:mag}\,(a). The fast saturating behaviour of $M_\mathrm{NC}$, occuring for all samples for magnetic fields $H\geq10$\,kOe, is characteristic of phase-separated Ga$\delta$FeN and resembles a Langevin function\,\cite{Navarro:2020_materials}. However, the presence of a weak magnetic hysteresis indicates that the majority of the NCs is not in thermal equilibrium. 

The temperature dependence of the normalized $M_{\mathrm{NC}}$ is shown in Fig.\,\ref{fig:mag}(b). The lines correspond to Langevin $M(T)$ fits, from where the $T_\mathrm{C}$ of the NCs is extrapolated, yielding values between (450$\pm$10)\,K and (500$\pm$10)\,K for the different samples. As an effect of the inclusion of Ga into the crystal lattice, these $T_\mathrm{c}$ values are lower than the 716\,K\,\cite{Dirba:2015_JMMM} and 767\,K\,\cite{Leinweber:1999_JAlloyComp} reported for $\gamma$'-Fe$_4$N. However, the $T_\mathrm{C}$, as well as the saturation value of $M_\mathrm{NC}^{\mathrm{sat}}$ at 300\,K, decrease with $y$, as shown in Fig.\,\ref{fig:mag}(c).
According to Burghaus \textit{et al.}, the magnetic moment of $\gamma$'-Fe$_4$N diminishes with the inclusion of Ga into 1$a$ sites. In contrast, in the nanocrystals considered here, where Ga occupies the 3$c$ sites in the lattice, the magnetization increases with increasing Ga. It is worth to underline, that the values of $M_\mathrm{NC}^{\mathrm{sat}}$ and $T_\mathrm{C}$ reported here originate from all the NCs in the samples, including those where no Ga is incorporated. 

The above results show, that the ferromagnetic character and magnetization values of the $\gamma$'-Ga$^{3c}_{y}$Fe$_{4-y}$N NCs embedded in GaN are stable up to 30\% of Ga in the lattice. This suggests, that the inclusion of Ga in 3$c$ sites does not affect significantly the magnetic properties of the system, in contrast to the case of Ga in 1$a$ sites reported for $\gamma$'-Ga$^{1a}_{y}$Fe$_{4-y}$N powders\,\cite{Burghaus:2010_IC,Burghaus:2011_JSSC}.

\section{Conclusion}
It has been shown that the combination of ADX and DAFS is a powerful tool to address the amount and local structure of Ga in fcc $\gamma$’-Ga$_y$Fe$_{4-y}$N nanocrystals embedded in a wurtzite GaN host. The inclusion of Ga into the embedded NCs has been confirmed for all samples and the amount of Ga in the NCs correlates with the normalized Ga-flow employed during the fabrication of the layers: the lower the Ga-flow provided, the less Ga is included in the NCs. A detailed analysis of the FT of the oscillations in the DAFS spectra shows, that Ga preferentially occupies the 3$c$ Wyckoff sites in the $\gamma$’-Ga$_y$Fe$_{4-y}$N lattice, in contrast to previously reported results for the same material system in the powder phase\,\cite{Houben:2009_ChemMat, Burghaus:2010_IC, Burghaus:2011_JSSC}. This is attributed to the arrangement of the surrounding GaN matrix, where Ga tends to occupy lattice sites with N as nearest neighbor, \textit{i.e.} occupying the 3$c$ sites with a reduced Ga-N distance.

The normalized magnetization of the $\gamma$’-Ga$^{3c}_y$Fe$_{4-y}$N nanocrystals embedded in GaN shows an inverse dependence with the amount $y$ of Ga obtained from the DAFS measurements: the more Ga in the NCs, the higher the magnetization. The same behaviour is observed for the $T_\mathrm{C}$ of the NCs, which lies in the range between 450\,K and 500\,K. These results show, that the $\gamma$’-Ga$^{3c}_y$Fe$_{4-y}$N NCs embedded in GaN are ferromagnetic with $T_\mathrm{C}$ values well above room temperature up to a Ga content of 30\% cations, suggesting that the incorporation of Ga into the 3$c$ sites does not affect significantly the magnetic properties of the system.

The present results shed light onto the long-standing issue of the unintentional incorporation of Ga atoms into $\gamma$’-Fe$_4$N nanocrystals embedded in a GaN matrix. The challenge of distinguishing between the Ga atoms in the host matrix and the ones in the embedded NCs -- a task beyond conventional element-sensitive spectroscopies -- has been here overcome with the combination of ADX and DAFS at the Ga $K$-edge, with an approach that can be extended to a large variety of magnetic phase-separated material systems\,\cite{Dietl:2015_RMP}.

\section{Experimental Section}
\label{exp}
\subsection{Sample fabrication}
The investigated samples have been fabricated by metalorganic vapor phase epitaxy by employing trimethyl-gallium (TMGa), ammonia and cyclopentadienyl-iron (CpFe$_2$) as sources for Ga, N and Fe, respectively. The TMGa source is kept at 0$^\circ$C and the CpFe$_2$ source at 17$^\circ$C. The array of NCs is grown on a 1\,$\mu$m thick GaN buffer layer in a $\delta$-growth mode, consisting of alternated growth periods of GaN:Fe and FeN\,\cite{Navarro:2019_crystals, Navarro:2012_APL}, on a $c$-plane sapphire (Al$_2$O$_3$) substrate. A 30\,nm GaN nucleation layer is deposited to compensate the lattice mismatch between substrate and buffer.

\subsection{Characterizations}
The fabricated samples have been pre-characterized by 2$\theta$-$\omega$ radial scans taken along the [0001] growth direction by high-resolution x-ray diffraction (HRXRD) with a 8\,keV Cu $K$-$\alpha$ radiation source with 8\,keV in a Panalytical Material Research Diffractometer. The measurements are performed employing a hybrid monochromator with a 0.25$^\circ$ slit for the incoming beam and a 5\,mm slit in front of a PixCel detector working with 19 detection channels. From the 2$\theta$-$\omega$ radial scans, the NCs phase and lattice parameter along the growth direction are obtained.

Further structural characterization is carried out \textit{via} transmission electron microscopy (TEM) imaging using a JEOL JEM-2200FS TEM microscope operated at 200\,kV in high-resolution imaging (HRTEM) mode. The TEM specimens are prepared in cross-section and in plan-view by a conventional procedure including mechanical polishing followed by Ar$^+$ milling.

In order to quantify the amount of Ga and to obtain the local structure of the Ga atoms in the NCs, anomalous x-ray diffraction (AXD) and diffraction anomalous fine structure (DAFS) at the Ga $K$-edge have been carried out at the SIRIUS beamline of the SOLEIL Synchrotron\,\cite{Ciatto:2016_TSF} selecting the symmetric (002) diffraction plane, expected to contain both Fe and Ga atoms. The setup exploits a double crystal monochromator equipped with Si(111) crystals to scan the energy of the incident photons. In order to reject high-order harmonics and to focus the beam, Pt-coated mirrors are employed, and a high vacuum diffractometer\,\cite{Ciatto:2019_JSR} serves as end-station. The diffraction signal and fluorescence background are collected by using a high vacuum PILATUS3 100K-M two-dimensional detector provided by Dectris.

\emph{Ab initio} DAFS simulations in the proximity of the Ga $K$ absorption edge have been performed using the FDMNES code\,\cite{Bunau:2009_JPCM} in dipolar approximation within a non-muffin-tin finite difference method (FDM) approach\,\cite{Smolentsev:2006_RPC}; simulations are based on clusters with the atoms occupying 3$c$ and 1$a$ sites. Quantitative analysis and fitting of the Ga $K$-edge DAFS oscillations, after correction of the DAFS amplitudes and phase shifts \cite{Prioetti:1999_PRB}, are carried out by using the FEFF package\,\cite{Newville:2001_JSR}, assuming the occupation of 3$c$ sites by Ga.

The magnetic properties are investigated in a Quantum Design superconducting quantum interference device (SQUID) MPMS-XL magnetometer equipped with a low field option at magnetic fields $H$ up to 70\,kOe in the temperature range between 2\,K and 400\,K following an experimental protocol for small signals elaborated to eliminate artifacts and to overcome limitations associated with integral SQUID magnetometry\,\cite{Gas:2019_MST, Sawicki:2011_SST}.

\section*{Conflicts of interest}
The authors declare no conflicts of interest.

\section*{Acknowledgements}
The authors acknowledge financial support by the Austrian Science Fund FWF project numbers V-478 and P31423 and by the CALIPSO project of the European Union. The experiment at the SIRIUS beamline benefited from the SOLEIL beam time allocation No. 20180261.





%
\end{document}